\documentclass[USenglish,twocolumn]{article}

\ifx\directlua\undefined\ifx\XeTeXcharclass\undefined
  \usepackage[utf8]{inputenc}                           
  \else\RequirePackage[no-math]{fontspec}[2017/03/31]\fi 
  \else\RequirePackage[no-math]{fontspec}[2017/03/31]\fi 
\usepackage[sort&compress,square,numbers]{natbib}
\usepackage[big,online]{dgruyter}
\usepackage{epsfig,graphicx,tabularx}
\usepackage{graphicx}
\usepackage{graphics}
\usepackage{epsfig}
\usepackage{amsmath,amssymb,mathrsfs}
\usepackage{amsfonts}
\usepackage{amssymb}
\usepackage{verbatim,color,ulem}
\usepackage{physics}
\usepackage{marginnote}
\usepackage{rotating}
\usepackage{lipsum}

\newcommand{\mnl}[1]{}
\newcommand{\mnr}[1]{}

\newcommand{\beq}{\begin{equation}} \newcommand{\eeq}{\end{equation}}
\newcommand{\beqa}{\begin{eqnarray}} \newcommand{\eeqa}{\end{eqnarray}}

\newcommand{\cblue}{\color{black}} 

\usepackage{xcolor}
\definecolor{darkgreen}{rgb}{0.0, 0.5, 0.0}
\newcommand{\cgreen}{\color{black}}

\theoremstyle{dgthm}

\theoremstyle{dgdef}

\begin{document}

\articletype{Research Article}

\author*[1]{Francesco Lorenzi}
\author[2]{Luca Salasnich}
\affil[1]{Dipartimento di Fisica e Astronomia “Galileo Galilei” of the Università di Padova
and INFN Sezione di Padova, Via Marzolo 8, 35131, Padova, Italy. email: \textit{francesco.lorenzi.2@phd.unipd.it}.}
\affil[2]{Dipartimento di Fisica e Astronomia “Galileo Galilei” of the Università di Padova 
and INFN Sezione di Padova, Via Marzolo 8, 35131, Padova, Italy; Padua Quantum Technologies 
Research Center, Via Gradenigo 6, Padova, Italy; Istituto Nazionale di Ottica del Consiglio 
Nazionale delle Ricerche, Via Nello Carrara 2, Sesto Fiorentino, Italy.}
\title{Variational approach to multimode nonlinear optical fibers}
\runningtitle{Variational approach to multimode nonlinear optical fibers}

\abstract{We analyze the spatiotemporal solitary waves of a graded-index multimode optical 
fiber with a parabolic transverse index profile. Using the nonpolynomial Schr\"odinger equation 
approach, we derive an effective one-dimensional Lagrangian associated with the 
Laguerre-Gauss modes with a generic radial mode number $p$ and azimuthal index $m$. We show that
the form of the equations of motion for any Laguerre-Gauss
mode is particularly simple, and we derive the critical power for the collapse for every mode. By solving
the nonpolynomial Schr\"odinger equation, we provide a comparison of the stationary 
mode profiles in the radial and temporal coordinates. 
}
\journalname{Nanophotonics}
\journalyear{2025}
\journalvolume{}

\maketitle
\vspace*{-6pt}

\section{Introduction} 
\vspace*{-3pt}
Solitary waves are nonlinear structures that appear in many
different physical systems \cite{kivsharintegrable, kartashov}. In Bose-Einstein
condensates, they correspond to solutions of the Gross-Pitaevskii equation (GPE) that are stationary in time
\cite{pethick}, while in nonlinear fiber optics, they are solutions of the propagation equation, that is a nonlinear Schr\"odinger equation (NLSE),
that propagate in space with a fixed temporal shape \cite{agrawal-book}. {\cgreen The two equations are equivalent from a mathematical point of view \cite{sulem-sulem}. Solitonic solutions are also ubiquitous in other nonlinear models \cite{malomed, seadawy1, seadawy2, seadawy3, seadawy4, seadawy5}, and the research on solitons and solitary waves of nonlinear partial differential equations is a lively field of research, especially regarding multidimensional solitons \cite{malomedmulti1,
malomedmulti2}.}

\mnl{stability}
Solitons {\cgreen of the NLSE} are known to be unstable in
any spatial dimension greater than one in the case of zero external potential,
while, under some conditions in trap geometry or, in the optical case, the
index profile, they can be stabilized. In particular, {\cgreen this} has been shown both in {\cgreen fully} three-dimensional traps \cite{mazzarella, gammal} and
waveguide-like traps \cite{sala2002, sala2002condensate, poletti_chapter}, i.e. traps
where the potential along one of the dimensions is absent.

In multimode fibers, solitons have been studied {\cblue within different theoretical approaches \cite{karlsson, yu_old}, and have recently been observed \cite{observation}. The most popular and effective analytical techniques are coupled-mode theory \cite{poletti} and variational analysis with a discrete set of variational parameters \cite{gaeta2000, sun, antikainen}. Using these methods, effective propagation equations have been derived and simulated, and several phenomena related to short pulse propagation have been described, such as nonlinear modulation of transverse width and collapse instability \cite{sun}.}
A remarkable phenomenon that occurs in multimode fibers when power levels are
sufficiently high is the so-called catastrophic collapse instability, marked by
a transfer of power to small spatial scales, where higher-order effect
becomes non-negligible and typically disrupts the integrity of the materials
\cite{gaeta2000, brabec, ranka, coexistence2018}. 
\mnr{other topics}
The dynamics of multimode nonlinear optical
fields with a high number of modes have been considered using
thermodynamic-inspired arguments, and the predicted mode occupation for long
propagation distances have been shown to be compatible with experimental
realizations \cite{zitelli_attractor, zitelli, zhong}.

\mnr{parabolic GRIN}
Due to the availability of new few-mode fibers and
the technological perspective of spatial division multiplexing, the topic gained interest{\cgreen, and in recent times a variety of research works have been devoted to the topic \cite{min}. }
The present work is focused on a {\cgreen fundamental kind of multimode fiber, that is} graded index (GRIN) fiber{\cgreen, consisting of a} parabolic index profile \cite{ahsan}. {\cgreen This kind of fiber has been used to understand the properties of a generic multimode nonlinear optical system \cite{marcuse-book}.}
{\cgreen In the absence of nonlinearity, this fiber has been shown to be amenable to a simple analytical treatment by WKBJ, and the exact modes}  are
given by the Laguerre-Gauss modes. These modes {\cgreen are of interest in the study of the light orbital angular momentum \cite{karimi}, and} in full
analogy to the two-dimensional harmonic oscillator problem in quantum mechanics, {\cgreen they}
are described by two indices, a principal one and an azimuthal one. A special
combination of principal and azimuthal indices is known as the radial index and is related to the number of maxima present in the radial function of
the mode \cite{krenn, sa}. 
\mnr{NPSE}
Assuming a fixed light polarization \cite{poletti_chapter}, the problem of
propagation in the medium of a solitary wave has strong similarities with the 3D
cigar-shaped Bose-Einstein condensate \cite{sala2002}.  In this context, a special variational ansatz has been shown to lead to the so-called nonpolynomial Schr\"odinger equation (NPSE) for the quasi-one-dimensional scenario of atomic condensate with tight transverse confinement. This equation has been shown to be particularly effective in accurately predicting the shape of bright matter-wave solitons, their collective modes, and the mean-field collapse \cite{sala2002, sala2002condensate}. Moreover, it has been used to predict the dynamics of matter-wave solitons in waveguides and their interactions with other solitons or barriers \cite{cuevas, lorenziatomic}. The nonlinear term appearing in the Gross-Pitaevskii equation for Bose-Einstein condensates is related to the s-wave scattering length and, in the low-energy regime, corresponds to a local interaction \cite{pethick}. In nonlinear optical media, the nonlinearity is given by the Kerr effect, which for most situations of interest is considered instantaneous.

\mnl{our contribution}
In this work, we use the variational framework for the 3+1 NLSE for parabolic GRIN optical fibers.
We assume a quite general variational ansatz constituted by Laguerre-Gauss modes with variable transverse width. The states are indexed by an azimuthal mode index, indicated by $m$, and a radial mode index, indicated by $p$, corresponding to the number of maxima of the radial function minus one. For each mode, we perform the dimensional reduction of the original Lagrangian, obtaining an effective one-dimensional Lagrangian for the axial field and the transverse width. 
Unlike previous works, we keep the axial degree of
freedom as a function{\cgreen. This allows us to obtain more general results with respect to assuming a parametrized shape, that is a method seldom utilized in previous approaches \cite{karlsson, Wabnitz2023, sa}}.
Our results show that the equations of motion {\cgreen for each choice of transverse mode} are related by a simple transformation involving the mode indices. This leads to a common criterion for stability against collapse.
We also provide a simplified treatment of the NPSE by
computing its cubic-quintic approximation. {\cgreen From a numerical perspective, t}his kind
of equation is even faster to simulate than the NPSE, and unlike the usual cubic NLSE, it can capture the wave collapse \cite{yang2023}.

\mnl{we believe that...}
In the case of a multimode field, this analysis can shed light
on the topic of the stability of Laguerre-Gauss modes for high-power light in a GRIN fiber.
It may also serve for future efforts to analyze the nonlinear interaction of fields in space division multiplexing applications.

\section{Derivation of the 3+1 NLSE for optical fibers}

Let us consider the electromagnetic field within an isotropic inhomogeneous
medium, denoting the position vector as ${\bf r}=(x,y,z)$ and the angular
frequency as $\omega$. The d'Alembert equation of the electric field ${\tilde
{\bf E}}({\bf r},\omega)$ in the space-frequency domain is given by
\cite{agrawal-book}
\beq
\left[ \nabla^2 + \beta^2(\omega ) \right] {\tilde {\bf E}}({\bf r},\omega) =
{\bf 0} \; , 
\label{basic}
\eeq
where $c$ is the speed of light in vacuum,
$\nabla^2=\partial_x^2+\partial_y^2+\partial_z^2$ is the Laplace operator, and 
\beq 
\beta^2(\omega) = \epsilon_r(\omega) \mu_r(\omega) {\omega^2\over c^2} 
\eeq
with $\epsilon_r(\omega)$ the relative electric permittivity function and
$\mu_r(\omega)$ the relative magnetic permeability function. In our specific
case $\mu_r(\omega)=1$. 

We now consider a waveguide structure, i.e. we posit that the electric field
factorizes as follows 
\beq 
{\tilde {\bf E}}({\bf r},\omega) = \frac{1}{2}{\tilde \Phi}({\bf r},\omega ) \, e^{i
\beta_0 z} \, {\bf u} + \text{c.c.} 
\label{factor-L}
\eeq
where $\beta_0$ is a propagation wavenumber along the $z$ axis and ${\bf u}$ is
a fixed unit vector along the radius of the fiber \cite{poletti_chapter}. In this way we get 
\beq
i \beta_0 \partial_z {\tilde \Phi}({\bf r},\omega) = \left[ - {1\over
2}\nabla_{\bot}^2 + {1\over 2} \beta_0^2 - {1\over 2} \beta^2(\omega) \right]
{\tilde \Phi}({\bf r},\omega)  \; ,   
\label{basic-LL}
\eeq
neglecting the term $\partial_z^2{\tilde \Phi}$ under the assumption that
${\tilde \Phi}({\bf r},\omega)$ is slowly varying function of $z$ (paraxial
approximation). Here we utilize the transverse Laplacian as $\nabla_{\bot}^2=\partial_x^2+\partial_y^2$. 

In nonlinear Kerr media, $\epsilon(\omega)$ depends on the electric field. We
model the function $\beta^2(\omega)$ as follows
\beq
\beta^2(\omega) = \delta \, \omega^2 - 2 W(x,y) - 2 g \int {\tilde {\bf
E}}^*({\bf r},\omega') \cdot {\tilde {\bf E}}({\bf r},\omega-\omega') \ d\omega' 
\label{simple}
\eeq
where $\delta$ and $g$ are phenomenological constants, while $W(x,y)$ models a transverse confining potential of the nonlinear optical fiber, i.e. the transverse profile of the refractive index. We will assume that such a term represents a harmonic potential in the radial variable, with a characteristic length of $\ell_\perp>0$. In doing so, we assume to have a parabolic GRIN fiber. The parameter
$\delta$ models linear dispersion, whereas the constant $g$ models
the Kerr effect.
Inserting Eq. (\ref{simple}) into Eq. (\ref{basic-LL})  
and then performing the anti-Fourier transform from the frequency domain
$\omega$ to the time domain $t$ of the resulting equation, we obtain 
\beqa
i \beta_0 \partial_z \Phi({\bf r},t) = \Big[ - {1\over 2}\nabla_{\bot}^2 +
{\delta\over 2} \partial_t^2 + {1\over 2} \beta_0^2 + W(x,y)  \nonumber \\+ g
|\Phi({\bf r},t)|^2 \Big] \Phi({\bf r},t)  \; . 
\label{basic-LLL}
\eeqa
This is a $3+1$ nonlinear Schr\"odinger equation (NLSE) with a cubic
nonlinearity. Comparing this equation with the NLSE of quantum mechanics, there
is an exchange of the axial coordinate $z$ with the time coordinate $t$. 
In this work, we will focus only on the anomalous dispersion regime, namely the
case of $\delta$<0. From now on, we consider the axial coordinate $z$ in units of
$\beta_0^{-1}$, the radial coordinate $r$ in units of $\ell_\perp$, and the time coordinate $t$ in units of $|\delta|^{-1/2}$. In
this way, from Eq. (\ref{basic-LLL}) we obtain the following adimensional $3+1$
NLSE 
\beqa
i \partial_z \Phi(x,y,z,t) = \Big[ - {1\over 2}(\partial_x^2 + \partial_y^2 +
\partial_t^2)  
+ W(x,y)  \nonumber \\
+ g |\Phi(x,y,z,t)|^2 \Big] \Phi(x,y,z,t)  \; 
\label{basic-LLLL}
\eeqa
removing the constant $1/2$ which does not affect the dynamics. 

As discussed also in Ref. \cite{Wabnitz2023}, Eq. (\ref{basic-LLLL}) can be
interpreted as the Euler-Lagrange equation obtained by extremizing the action
functional 
\beq
S[\Phi] = \int L \, dt \;  ,
\eeq
with Lagrangian 
\beq 
L = \int \mathscr{L} \, dx \, dy \, dz \; ,
\eeq
and Lagrangian density 
\beqa 
&\mathscr{L} = \tfrac{i}{2} ( \Phi^* \partial_z \Phi - \Phi \partial_z \Phi^*)  
- {1\over 2}( |\partial_x\Phi|^2 +|\partial_y\Phi|^2 \nonumber \\
&+|\partial_t\Phi|^2 )- W(x,y) |\Phi|^2 - \frac{g}{2} |\Phi|^4 \; ,
\label{ladensity}
\eeqa
Equivalently, in cylindrical coordinates,
\begin{align}
&\mathscr{L} = \tfrac{i}{2} ( \Phi^* \partial_z \Phi - \Phi \partial_z \Phi^*)  
- \tfrac{1}{2}( |\partial_r\Phi|^2 \nonumber \\
&+\tfrac{1}{r^2}|\partial_\theta\Phi|^2 +|\partial_t\Phi|^2 ) - W(x,y) |\Phi|^2 
- \tfrac{g}{2} |\Phi|^4 \; .
\label{ladensitypolare}
\end{align}
So, there is a complete analogy with the Gross-Pitaevskii theory that describes
the mean-field properties of a Bose-Einstein condensate \cite{pethick}. In
particular, when considering the case of anomalous dispersion, the Lagrangian
density corresponds to the one of an attractive condensate.

\section{Nonpolynomial effective action}

\subsection{Laguerre-Gauss modes}
It is useful to work in the cylindrical reference frame to compute the Laguerre-Gauss modes. We select a family of
wavefunctions that correspond to such modes and are therefore indexed
by two integer numbers $n$ and $m$. They span the range $n = 0,1, 2,...$, and $m=-n, ..., n$. We will assume to have a space and time-dependent transverse scaling function $\sigma_{nS}(z, t)$ that regulates the width of the mode in the $x-y$ plane.  For brevity, we define $S=|m|$
\begin{equation}
\Phi_{nm}(r,\theta, z, t) = A_{nS}(z,t) \, T_{nS}(r;\sigma_{n S}(z,t)) \ e^{i m \theta},
\end{equation}
where the transverse wavefunctions are the eigenstates of the 2D harmonic
problem that naturally arises in the cylindrical coordinate frame,
\begin{align}
&T_{nS}(r, \sigma_{n S}(z,t))=\sqrt{\frac{p!}{\pi\sigma_{n S}^2(z,t)(p+S)!}} \nonumber \\ & \times \left(\frac{r}{\sigma_{n S}(z,t)}\right)^{S} \exp\left[-\frac{r^2}{2\sigma_{n S}^2(z,t)}\right]  L_p^{S}\left(\frac{r^2}{\sigma_{n S}^2(z,t)}\right),
\end{align}
where $n$ is the principal integer number, $m$ is the angular integer number,
and $p = (n-S)/2$ is the radial number. We start by analyzing the choice of
transverse variational \textit{ans\"atze} of the form \cite{salamalomed}
\begin{equation}\label{ansatzvecchia}
  T_{SS}(r, \sigma_{SS}(z, t)) = \frac{r^S}{\sqrt{\pi S!}  \ \sigma_{SS}^{S+1}(z, t)} \exp
  \left[-\frac{r^2}{2 \sigma_{SS}^2(z, t)}\right] 
\end{equation}
which are the Laguerre-Gauss modes under the condition of a nodeless radial
profile. This is reminiscent of the treatment of vortices in Bose-Einstein
condensates, where the condensate density near the origin follows a power law
with an exponent equal to the vorticity. In the lowest eigenstate, corresponding
to $n=0$, $m=0$, the generalized Laguerre polynomial is constant and equal to
$1$, so the state corresponds to the Gaussian variational state that was used in
\cite{sala2002} for the derivation of the NPSE.
The ansatz~(\ref{ansatzvecchia}) corresponds to Laguerre-Gauss states in which
$n=S$, or $p=0$, indicating a single maximum of the transverse function. This state will correspond to a simple vortical state with vorticity $m$. 

Following Salasnich, Malomed, and Toigo \cite{salamalomed}, we remark that, if
we choose an initial state with a null radial number $p=0$,
upon substitution and integration, we obtain
\begin{align}\label{p0}
&\mathscr{L}^{(0)}=  i A^* \partial_z A -\frac{1}{2}\left|\partial_t A\right|^2-\frac{1}{2}(S+1)\left(\frac{1}{\sigma^2}+\sigma^2\right)|A|^2  \nonumber \\
&-\frac{(S+1)}{2 \sigma^2}\left(\partial_t \sigma\right)^2|A|^2-\frac{1}{2}(S+1) g_{0S} \frac{|A|^4}{\sigma^2} \; ,
\end{align}
where we dropped the subscripts of the fields $A$ and $\sigma$, and we defined
\begin{equation}
    g_{0S} = \frac{g}{2\pi} \frac{(2S)!}{2^{2S}(S!)^2 (S+1)}.
\end{equation}
We can also choose a radial mode number that is different from zero. The results with $p=1, 2$ bear expressions similar to those found for $p=0$. In the case of a single-node state, the effective action is
\begin{align}\label{p1}
&\mathscr{L}^{(1)}=  i A^* \partial_z A -\frac{1}{2}\left|\partial_t A\right|^2-\frac{1}{2}(S+3)\left(\frac{1}{\sigma^2}+\sigma^2\right)|A|^2  \nonumber \\
&-\frac{(3S+5)}{2 \sigma^2}\left(\partial_t \sigma\right)^2|A|^2-\frac{1}{2}(S+3) g_{1S} \frac{|A|^4}{\sigma^2} \; .
\end{align}
and with 
\begin{equation}
    g_{1S} = \frac{g}{2\pi} \frac{(3S+2)\Gamma(\frac{1}{2}+S)}{4 \pi^{1/2} (S+1)!(S+3)}.
\end{equation}
For the $p=2$ case, we have
\begin{align}\label{p2}
&\mathscr{L}^{(2)}=  i A^* \partial_z A -\frac{1}{2}\left|\partial_t A\right|^2-\frac{1}{2}(S+5)\left(\frac{1}{\sigma^2}+\sigma^2\right)|A|^2  \nonumber \\
&-\frac{(5S+13)}{2 \sigma^2}\left(\partial_t \sigma\right)^2|A|^2-\frac{1}{2} (S+5)g_{2S} \frac{|A|^4}{\sigma^2} \; .
\end{align}
and with 
\begin{equation}
    g_{2S} = \frac{g}{2\pi} \frac{(44+95S+41S^2)(2S)!(S+1)(S+2)}{2^{2S+6} ((S+2)!)^2(S+5)}.
\end{equation}

We derive a general result for every index $p$ and $S$ by following the procedure described in Appendix A. In the following, we neglect the term containing the derivatives of the transverse width $\sigma$, that are reported in Appendix A. Therefore, we can write the effective Lagrangian density as
\begin{align}\label{NPSELagrangian}
     &\mathscr{L}^{(p)}=  iA^* \partial_z A - \frac{1}{2}\left|\partial_t A\right|^2
     \nonumber \\ &-\frac{1}{2}\xi_{pS}\left(\left(\frac{1}{\sigma^2}+\sigma^2\right)|A|^2 + g_{pS} \frac{|A|^4}{\sigma^2} \right)\; .
\end{align}
with $\xi_{pS} = S+2p+1$, and 
\begin{align}\label{gpS}
    &g_{pS} = \frac{g}{2\pi} \frac{1}{\xi_{pS}2^{4p+2S}} \nonumber \\& \times \sum_{q=0}^{p}\frac{(2q)![(2p-2q)!]^2(2S+2q)!}{(q!)^2[(p-q)!]^4[(S+q)!]^2}.
\end{align}
We can relate this class of effective
Lagrangians to the usual NPSE Lagrangian derived by assuming the fundamental
mode \cite{sala2002}. This is done by using a change of variables, namely $z \to
z/\sqrt{\xi_{pS}}$, $t \to t/\xi_{pS}$, $g_S^{(p)}\to g$. The corresponding
Euler-Lagrange equations consist of a (1+1)-dimensional PDE for the axial field
$A$ and an algebraic equation for the variational width $\sigma$. 
\begin{equation}\label{EL1}
    i\partial_z A = -\frac{1}{2}\partial_t^2A + \xi_{pS}\left(\frac{1+\sigma^4 + 2g_{pS}|A|^2}{2\sigma^2} \right)A,
\end{equation}
and 
\begin{equation}
    \sigma^4 = 1+g_{pS}|A|^2,
\end{equation}
Since the Euler-Lagrange equation for the function $\sigma$ is indeed algebraic,
we can substitute the value of $\sigma$ back into Eq.~(\ref{EL1}), to obtain the
NPSE for the Laguerre-Gauss modes,
\begin{equation}\label{NPSE}
    i\partial_z A = -\frac{1}{2}\partial_t^2A + \xi_{pS}\left(\frac{1 + (3/2)g_{pS}|A|^2}{\sqrt{1+g_{pS}|A|^2}} \right)A,
\end{equation}
Notice that this expression includes the one obtained in Ref.~\cite{malomed} as
a special case for $p=0$, being $\xi_{0S} = (S+1)$.

\subsection{Solitonic solutions and collapse}
\begin{figure}
    \centering
    \includegraphics[width=0.95\linewidth]{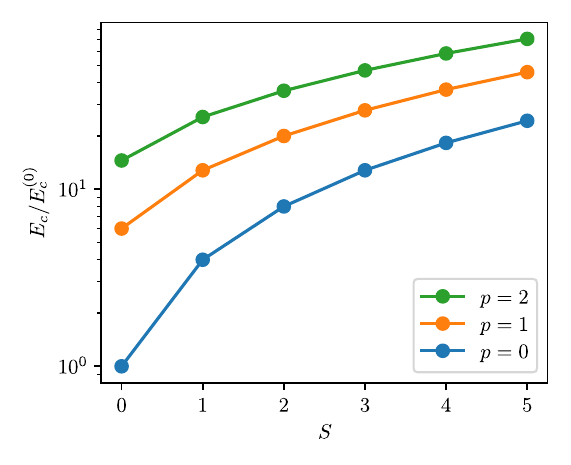}
    \caption{Critical pulse energy as a function of the mode numbers.}
    \label{critical}
\end{figure}
The shape of the solitonic solution can be derived in analogy to
Ref.~\cite{sala2002}. We utilize the stationary ansatz $A(z, t) = a(t)\exp[-i
\kappa z]$. 
Furthermore, we define the normalized interaction parameter
$\gamma_{pS} = |g_{pS}|$, and a rescaled propagation variable $\kappa_{pS}= \kappa/\xi_{pS}$. Substituting into Eq.~(\ref{NPSE}), we 
{\cblue obtain the stationary equation
\begin{equation}\label{stationary-npse}
    \kappa a = -\frac{1}{2}\partial_t^2a + \xi_{pS}\left(\frac{1 + (3/2)g_{pS}|a|^2}{\sqrt{1+g_{pS}|a|^2}} \right)a \, .
\end{equation}
It }can determine
the temporal shape of the soliton via quadrature. Indeed, by imposing that the soliton amplitude vanishes for $t \to \pm \infty$, we obtain the implicit relation
\begin{align}
    &t = \frac{1}{\sqrt{2 \ \xi_{pS}}} \nonumber \\ 
    \times\Bigg[&\frac{1}{\sqrt{1-\kappa_{pS}}}    \arctan\left(
      \sqrt{\frac{\sqrt{1-\gamma_{pS}a^2} - \kappa_{pS}}{1-\kappa_{pS}}}
    \right) \nonumber \\
    &- \frac{1}{\sqrt{1+\kappa_{pS}}}  \operatorname{arctanh}\left(
     \sqrt{\frac{\sqrt{1-\gamma_{pS}a^2}-\kappa_{pS}}{1+\kappa_{pS}}}
    \right) \Bigg].
\end{align}
\begin{figure}[h]
    \centering
    \includegraphics[width=1\linewidth]{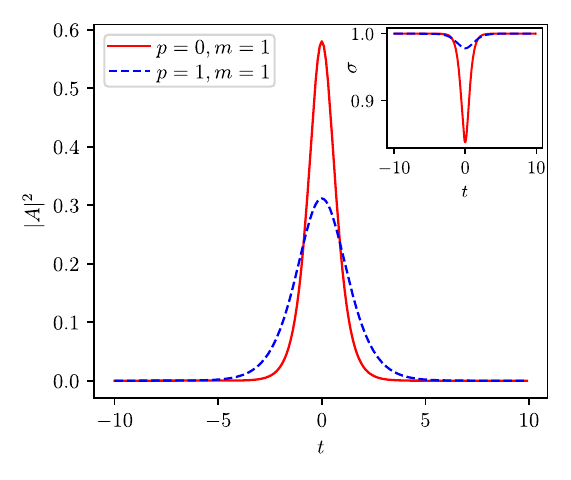}
    \caption{Temporal profile in time for two solitonic solutions {\cblue of the stationary NPSE Eq.~(\ref{stationary-npse}),} corresponding to the $m=1$ states with $p=0$ (solid red line) and $p=1$ (dashed blue lines). In the inset, the variational transverse width $\sigma$ over the same time span. The time $t$ is in units of $|\delta|^{-1/2}$.}
    \label{axial}
\end{figure}
We can impose the normalization condition for the pulse normalized energy $E = \int_{-\infty}^{\infty} dt |A(z, t)|^2$ therefore obtaining a relation between the 
propagation variable and the nonlinearity parameter 
\begin{equation}\label{Ek}
    E = \tfrac{2\sqrt{2}}{3\gamma_{pS}} (2\kappa_{pS}+1)\sqrt{1-\kappa_{pS}}.
\end{equation}

\begin{figure}
    \centering
    \includegraphics[width=\linewidth]{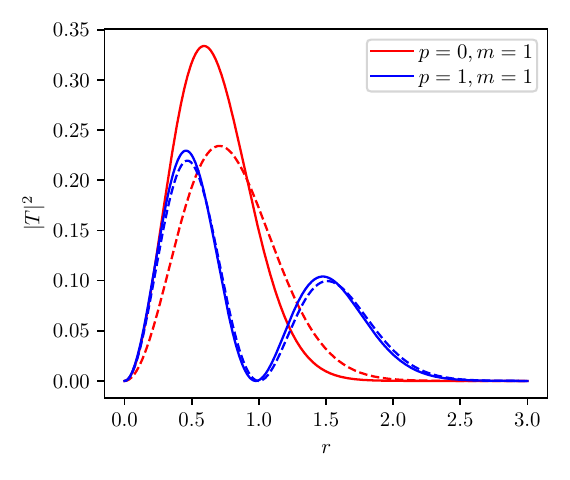}
    \caption{Radial profile for two solitonic solutions {\cblue of the stationary NPSE Eq.~(\ref{stationary-npse}),} corresponding to the $m=1$ states with $p=0$ (red lines) and $p=1$ (blue lines). The solid line is the radial shape at the time corresponding to the peak of the soliton, and the dashed line corresponds to the linear case, which approximates the low-energy regime. $r$ in units of $\ell_\perp$.}
    \label{radial}
\end{figure}

This relation has solutions for every $E < E_c$ that is the threshold energy for the collapse instability of the optical field. Moreover, this
point is the point of union of the stable and unstable branches as marked by the
Vakhitov-Kolokolov (VK) criterion. {\cblue The VK criterion is} a necessary condition for the
stability of the solitons, that is, the solitons are stable {\cblue only if} $\partial E/\partial\kappa_{pS}>0$. {\cblue The criterion is a general result for a stationary solution of conservative nonlinear partial differential equations \cite{vk}}. By computing the derivative of Eq.~(\ref{Ek}), we obtain this critical value as
\begin{equation}\label{critical-energy}
    E_c = \frac{4}{3\gamma_{pS}}.
\end{equation}
We plot the critical energies normalized with respect to $E_c^{(0)}=4/(3\gamma_{00})$ for some values of $p, S$ in Fig.~{\ref{critical}}.  
It follows that, with a fixed value of the Kerr coefficient $g$, modes with large indices $S$ and $p$ are more stable against collapse. {\cblue We solve the stationary NPSE for a couple of representative modes with the index $m=0$, and $p=0, \ 1$. These two modes correspond to a ring-like mode and a double ring-like one.}
{\cblue The solution is computed by} using a split-step Fourier method implemented in Ref.~\cite{lorenzi}, with the choice of parameter $g=-1$ and a normalized pulse energy of $E = 1$.
In Fig.~\ref{axial}, we plot the temporal profiles of solitons solved with the NPSE. In Fig.~\ref{radial}, we plot their respective radial profiles. The effect of nonlinearity on the field distribution in both temporal and radial coordinates is more pronounced for the mode with the lower radial number{\cblue, i.e. the mode with the higher nonlinear coefficient that with this choice of parameters amounts  to $\gamma_{01}=1/(8\pi)\approx 3.98 \times 10^{-2}$ versus $\gamma_{11}=5/(128\pi)\approx 1.24 \times 10^{-2}$.}

\section{Cubic-quintic approximation}
The results of the application of the VK criterion show that the NPSE, for pulse energies less than a critical value $E_c$, possesses two soliton branches, one stable and one
unstable. We can approximate the NPSE by means of an expansion of the nonpolynomial part in a power series. This, to the lowest order in the nonlinearity, consists of the usual cubic equation (indicated by subscript $3$), and to the next order to a cubic-quintic equation (indicated by subscript $3-5$). Their solitonic solutions are always stable with respect to the VK criterion, but the cubic-quintic, contrary to the cubic
equation, also possesses a maximum energy of the soliton \cite{sala2002}, which can also be taken as a critical energy, $E_c$. By integrating the stationary version of Eq.~(\ref{NPSE}), we obtain
\begin{equation}
    \frac{da}{dt}=\sqrt{-V(a)},
\end{equation}
with the effective potential being given by
\begin{equation}
    V(a) = -\kappa_{pS} a^2 - a^2\sqrt{1-\gamma_{pS}a^2}.
\end{equation}
The potential can also be substituted with power series expansions to quartic order
\begin{equation}
    V_{3}(a)= -a^2\left[(\kappa_{pS}+1)-\frac{1}{2} \gamma_{pS} a^2\right]
\end{equation}
for the cubic equation, and to sixtic order
\begin{equation}
    V_{3-5}(a)=-a^2\left[(\kappa_{pS}+1)-\frac{1}{2} \gamma_{pS} a^2-\frac{1}{8} \gamma_{pS}^2 a^4\right]
\end{equation}
for the cubic-quintic equation.
The pulse energy is calculated by means of a change in variables and
integrating over the soliton shape \cite{sala2002condensate}. In the cubic case, it is
\begin{equation}\label{3}
    E_{3}=\frac{2\sqrt{2}}{\gamma_{pS}}\sqrt{1+\kappa_{pS}} \, ,
\end{equation}
and in the cubic-quintic cases, it is
\begin{equation}
    E_{3-5} = \frac{4}{\gamma_{pS}} \Big(\frac{\pi}{2}- \arctan\left(\frac{\sqrt{2}\sqrt{1+\kappa_{pS}}}{\sqrt{3+2\kappa_{pS}}-1}\right)\Big) \, .
\end{equation}
The cubic-quintic pulse energy can be rewritten by using the properties in Appendix B, as
\begin{equation}\label{3-5}
    E_{3-5} = \frac{2}{\gamma_{pS}}\arctan(\sqrt{2}\sqrt{1+\kappa_{pS}}) \, .
\end{equation}
We remark that the cubic-quintic treatment gives a cubic interaction coefficient as in the case of the one reported in Eq. (10) of Ref. \cite{malomedk}, but it has a different value.
\begin{table}[h]
    \centering
    \begin{tabular}{c|c|c|c}
         & C & CC &  NPSE \\
        \hline
        $\partial E/\partial \kappa_{pS}$    & $>0$    & $>0$  & $\mathrm{2 \; branches}$   \\
        \hline
        $E_C $   & $+\infty$  & $\pi/\gamma_{pS}$    & $4/(3\gamma_{pS})$    \\
    \end{tabular}
    \caption{Stability with the Vakhitov-Kolokolov criterion and critical energy for different quasi-one dimensional effective equations. 
    }
    \label{tab2}
\end{table}

The VK criterion does not indicate a critical energy for the case of cubic and cubic-quintic approximations, since the energy functions (\ref{3}) and (\ref{3-5}) are monotonic functions of the propagation variable $\kappa_{pS}$. However, we notice that in the cubic-quintic case, there is an upper bound on the energy that is found by taking the limit of $\kappa_{pS}$ to infinity. This value corresponds to $E_{3-5 \, c} = \pi/\gamma_{pS}$, and can be considered as a critical value for the energy. On the other hand, no such upper bound exists for the cubic equation. We summarize the results of the application of the VK criterion and the maximum pulse energy in Tab.~\ref{tab2}. From these results, it is clear that the cubic equation is unable to predict the collapse, whereas the cubic-quintic indicates a collapse with an energy that is $\approx2.3$ times higher than the one indicated by the NPSE.

{\cgreen 
\section{Comparison with related works}
The method presented in this paper provides an alternative to other methods such as variational analyses with a discrete set of parameters \cite{sun, sa} and coupled-mode theory \cite{poletti}. In particular, it is more flexible in the description of axial field amplitudes that are difficult to parametrize, like the one resulting from modulation instability of CW radiation. The scaling relation Eq.~(\ref{gpS}) has been proven to be useful also in the case of nonlinear propagation of Laguerre-Gauss modes in bulk optics \cite{sa}, where also stability of mixed modes was addressed by means of a variational analysis in the beam parameters. Recent works focusing on optical fibers utilize a variational method within the discrete set of pulse parameters to determine stability in GRIN fibers \cite{Wabnitz2023}, determining stability using the Vakhitov-Kolokolov criterion, along with the Lyapunov criterion and evaluating the spectrum of linear perturbation at the solitonic fixed point. Our results on the critical energy of the pulse, taken from in Eq.~(\ref{critical-energy}), also reported in Fig.~(\ref{critical}), can be compared with the said work in the case of the fundamental spatiotemporal soliton. In this case the values provided in \cite[Eq.~(42)]{Wabnitz2023} overestimate slightly the value of the critical energy. Remarkably, the variational ansatz used in Ref.~\cite{Wabnitz2023} bear profound similarities to the one utilized in Ref.~\cite{sala2002condensate}, obtained in the context of atomic condensates, the only difference being that in the former case the axial field is assumed in a hyperbolic secant shape, whereas in the latter case it is assumed to be a Gaussian. Our results, obtained with a modified version of the NPSE, are instead more general, since the equation admits an arbitrary shape for the axial field amplitude.}

\section{Conclusions}
We have analyzed solitary waves related to Laguerre-Gauss modes in parabolic GRIN fibers by using a variational method in which the variational degrees of freedom are the axial field and its transverse width. We have obtained the effective one-dimensional Lagrangians for the cases of a radial index $p=0, 1, 2$ and arbitrary azimuthal index $S$. Our calculations have shown that the resulting equations of motion are formally very similar. We have developed a mapping resulting in a common description of the soliton stability region with the Vakhitov-Kolokolov criterion and compared our results with the previous characterization of~\cite{sa}. Finally, we proposed a cubic-quintic equation as an improved method to describe nonlinear propagation. Our results can also be applied to the case of quasi-one-dimensional Bose-Einstein condensates, where this treatment on Laguerre-Gauss modes has not yet been performed. {\cgreen We envisage future works to address the problem of finding the solitonic states resulting from a transverse profile composed of a mixture of Laguerre-Gauss modes and investigating their stability. A more application-oriented future direction of this work may investigate the effect of fiber core ellipticity (random or deliberate) and birefringence on the properties of higher-order solitons based on Laguerre-Gauss modes.}

\subsection*{Research funding}
LS acknowledges the BIRD Project 
“Ultra-cold atoms in curved geometries” of the University of
Padova. LS is partially supported by the European
Union-NextGenerationEU within the National Center
for HPC, Big Data, and Quantum Computing [Project
No. CN00000013, CN1 Spoke 10: Quantum Computing] and by the European Quantum Flagship Project
PASQuanS 2. LS acknowledges Iniziativa Specifica
Quantum of Istituto Nazionale di Fisica Nucleare, the
Project “Frontiere Quantistiche” within the 2023 funding programme ‘Dipartimenti di Eccellenza’ of the Italian
Ministry for Universities and Research, and the PRIN
11 2022 Project “Quantum Atomic Mixtures: Droplets,
Topological Structures, and Vortices”.

\subsection*{Competing interests}
Authors state no conflict of interest.

\section*{Appendix A}
In order to compute the generic reduced Lagrangian density for any $p$ and $S$, we follow the approach proposed in Ref.~\cite{sa}. However, unlike this approach, which focuses on unguided beams, we have a guided propagation of pulses, so in our original Lagrangian an additional term like $|\partial_t\Phi|^2$ is present, as well as the guiding term $W(x, y)$. The main relationship is the explicit calculation of the integrals of the kind 
\begin{align}
    I_{mnl} &= \int_{0}^{\infty} dr \, \frac{2r}{\sigma^2} \exp\left[-m\frac{r^2}{\sigma^2}\right] \left(\frac{r^2}{\sigma^2}\right)^{mS-l}\nonumber\\ 
    &\times \left[L_p^{S}\left(\frac{r^2}{\sigma^2}\right)\right]^{2m-l}\left[L_{p-1}^{S+1}\left(\frac{r^2}{\sigma^2}\right)\right]^{n}.
\end{align}
These integrals can be computed using the fundamental properties \cite{nist, abramowitz}
\begin{align}
&\int_0^{\infty} e^{-x} x^{|\ell|} L_p^{|\ell|}(x) L_q^{|\ell|}(x) d x  =\frac{(p+|\ell|)!}{p!} \delta_{p q}, \\
&\int_0^{\infty} e^{-x} x^{|\ell|+1}\left[L_p^{|\ell|}(x)\right]^2 d x =\frac{(p+|\ell|)!}{p!}(2 p+|l|+1), \\
&L_p^{|\ell|}(x) =L_p^{|\ell|+1}(x)-L_{p-1}^{|\ell|+1}(x), \\
&L_p^{|\ell|+1}(x) =\sum_{k=0}^p L_k^{|\ell|}(x) .
\end{align}
In addition to the terms derived in Ref.~\cite{sa}, we also derive the terms
\begin{align}
    &I_{122} = \frac{(p+S+1)!}{(p-1)!} + \frac{(p+S)!}{(p-2)!},\\
    &I_{112} = -2\frac{(p+S+1)!}{(p-1)!} - \frac{(p+S)!}{(p-2)!},\\
    &I_{102} = \frac{(p+S+2)!}{(p)!} + 4\frac{(p+S+1)!}{(p-1)!}+\frac{(p+S)!}{(p-2)!},
\end{align}
that are needed for the integration of the novel terms.
If the derivative of the $\sigma$ is not neglected, we will have in the Lagrangian density and additional term of the form 
\begin{align}
    & \mathscr{L}_{\partial_t\sigma}^{(p)} = -\frac{1}{2}|A|^2\left(\frac{\partial_t\sigma}{\sigma}\right)^2 \Big[\frac{p!}{2(p+S)!}\Big(4  I_{122} + \nonumber \\
    &4 I_{112} - 4 S  I_{111} + I_{102} -  2 S I_{101}\nonumber \\ &+  S^2 I_{100} - 2 I_{111} - I_{101} + S  I_{100} \Big)\Big]= \nonumber\\
    &=-\frac{1}{2}|A|^2\left(\frac{\partial_t\sigma}{\sigma}\right)^2 (1+S+2p(1+p+S)).
\end{align}
Substituting the values of $p=0, 1, 2$, the results in Eqs.(\ref{p0}), (\ref{p1}) and (\ref{p2}) are retrieved.

\section*{Appendix B}
As useful mathematical results, we recall the following relationships of the inverse tangent function. The expression for the cubic-quintic pulse energy can be
simplified by noting that
\begin{align}
    &\frac{1}{2}\arctan\left(\sqrt{2}\sqrt{1+\kappa_{pS}}\right) = \frac{\pi}{2} \nonumber \\ 
    &- \arctan\left(\frac{\sqrt{2}\sqrt{1+\kappa_{pS}}}{\sqrt{3+2\kappa_{pS}}-1}\right),
\end{align}
that is a special version of the basic relationship 
\begin{equation}
    \frac{1}{2}\arctan(x) = \arctan\left(\frac{\sqrt{1+x^2}-1}{x}\right),
\end{equation}
together with 
\begin{equation}
    \arctan(x)+\arctan(1/x) =\pi/2 \, .
\end{equation}
Both identities can be derived from the sum formula for the tangent, $\tan(x+y)=(\tan(x)+\tan(y))/(1-\tan(x)\tan(y))$. 


\newpage
\end{document}